\documentclass[%
 reprint,
showpacs,
 amsmath,amssymb,
 aps,
]{revtex4-1}

\usepackage{graphicx}
\usepackage{dcolumn}
\usepackage{bm}

\begin{document}

\title{Ghost Waves in Anisotropic Materials: negative refractive index and evanescent field enhancement in lossless media.}


\author{Evgenii Narimanov}
\affiliation{School of Electrical and Computer Engineering, and Birck Nanotechnology Center, Purdue University, West Lafayette IN 47907}

\date{\today}

\begin{abstract}
We show that transparent dielectrics with strong optical anisotropy support a new class of electromagnetic waves that combine the 
properties of propagating and evanescent fields. These ``ghost waves" are created in tangent bifurcations that ``annihilate'' pairs of  positive- and negative-index modes, and represent the optical analogue of the ``ghost orbits" in the quantum theory of non-integrable dynamical systems.\cite{Kus} Similarly to the regular evanescent fields, ghost waves support high transverse wavenumbers, but in addition to the exponential decay show oscillatory behavior in the direction of propagation. Ghost waves can be resonantly coupled to the incident evanescent waves, which then grow exponentially through the anisotropic media -- as in the case of negative index materials.\cite{Veselago,Pendry} As ghost waves are supported by transparent dielectric media, they are free from the ``curse'' of material loss that is inherent to conventional negative index composites.
\end{abstract}

\maketitle

The recent development of negative index metamaterials, where the subwavelength structure of the composite unit cell allows  simultaneously negative electric and magnetic response, \cite{Veselago} gave rise to the experimental demonstration of such unusual phenomena as negative refraction and backwards wave propagation \cite{NR-microwaves}, electromagnetic cloaking \cite{cloaking-microwaves} and subwavelength focusing. \cite{TL-Grbic} Particularly striking
is the behavior of the evanescent waves  which instead of the expected decay may show exponential growth through the negative index medium,\cite{Pendry} which offers the potential of to revolutionize the near-field optics.\cite{Zhaowei_APL} However, this effect is only present when the metamaterial unit cell size is much smaller than the wavelength in the medium. \cite{Roberto_PNAS} As a result, while there are many different realizations of the negative index media -- from photonic crystals \cite{NIPC-Joannopoulos} to coupled Mie resonators \cite{NI-Mie} to plasmonic composites \cite{Vlad-book},  it is only the latter class that supports this exponential enhancement of evanescent field. 

However, the material loss inherent to plasmonic media due to the inevitable free-carrier absorption\cite{ftn1} severely limits the evanescent field enhancement \cite{EN_OL} Despite multiple attempts to remove this stumbling block with new materials \cite{Sasha_AM}  or incorporating material gain in the composite design \cite{Vlad_Science}, the  (nearly) ``lossless metal'' \cite{Jacob_APL} that would allow the evanescent field control and amplification promised by metamaterial research for nearly two decades since the seminal work of J. Pendry, \cite{Pendry}  remains an elusive goal.\cite{Sasha_AM}

In this Letter, we present an alternative to the plasmonic approach. We demonstrate that strongly 
anisotropic dialectics  support ``ghost waves'' that differ from the ``regular" propagating and evanescent fields. These ghost waves represent the optical analogue of the ghost orbits in the semiclassical theory of non-integrable systems. Similarly to the surface plasmons in negative index media, these ghost waves can resonantly couple to
the incident evanescent field, resulting in its exponential ``amplification'' across the anisotropic media. However, 
all the primary component of the dielectric permittivity tensor for such the strongly anisotropic media that supports the ghost waves, can be positive. As a result, the effect can be observed in a transparent and (nearly) lossless dielectric.

\begin{figure}[htbp] 
   \centering
    \includegraphics[width=3.5in]{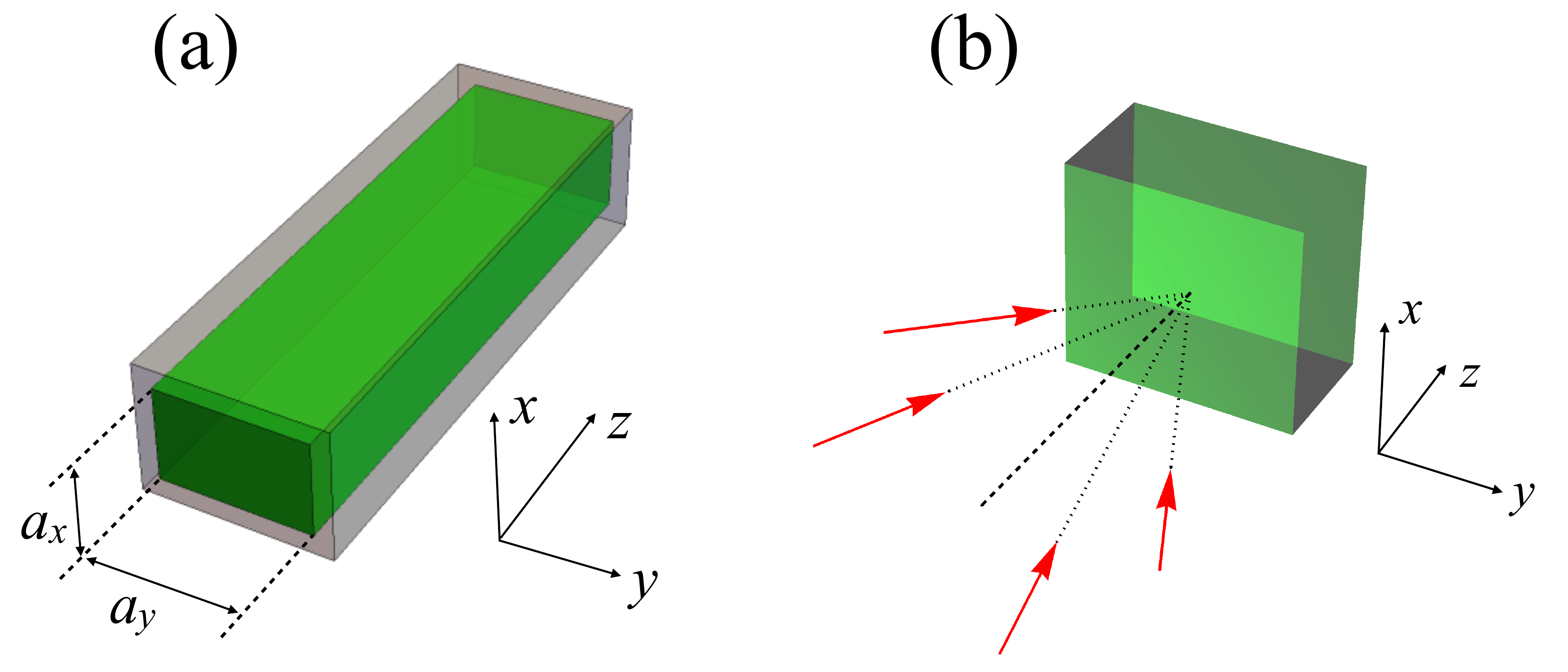}
     \caption{The schematics of the metal-clad waveguide with the core formed by a biaxial anisotropic dielectric material (a), and the corresponding standing-standing wave pattern formed by the interference of several beams incident onto a slab of the biaxial anisotropic dielectric (b). }     
     \label{fig:1}
\end{figure}

Due to the complexity of the fully three-dimensional nano-fabrication required for the metamaterials with simultaneously negative values of the dielectric permittivity and magnetic permeability, much attention was given 
to the possible alternatives that do not rely on the magnetic response.\cite{Christodoulides_OE,EN_OE,Moti_OE} If the desired negative index performance can be  limited to the propagation in a waveguide, a number of such solutions are possible -- using hyperbolic metamaterials, \cite{EN_OE} Clarricoats-Waldron geometry \cite{Christodoulides_OE}, or a waveguide with the core formed by biaxial anisotropic dielectric \cite{Moti_OE} -- see Fig. \ref{fig:1}(a). Note that in each of these solutions, the negative index modes appear in pairs with their positive index counterparts, with each pair born together at the same point of the system phase space at some critical frequency $\omega_c$.  

In direct proximity of $\omega_c$, the mode dispersion $\omega\left(k_z\right)$ can be treated as a bifurcation diagram with the the frequency taking the role of the control parameter. Known as the tangent bifurcation, it is subject to the 
general rules of nonlinear dynamics,\cite{NL_book} such as e.g. the creation of new states in integer number of pairs.  In particular, the actual mode spectra below the tangent bifurcation threshold show the signatures of the so called ``ghost orbits" formally defined as extensions of the system dynamics to a higher-dimensional phase space (e.g. extended to complex time and/or spatial coordinates).\cite{Kus} We therefore conclude that the negative index 
 systems in a waveguide geometry that do not rely on magnetic response, \cite{Christodoulides_OE,EN_OE,Moti_OE} actually support optical equivalents of these ``ghosts''. Furthermore, such optical ghost waves   are essentially different from both positive- and negative-index modes.
 
 While a waveguide with the core formed by a hyperbolic metamaterial, \cite{EN_OE} with suffer high propagation loss due to the absorption in the hyperbolic medium, the approach based on biaxial dielectric media  \cite{Moti_OE} is
 is not so limited. Even with the metallic cladding (see Fig. \ref{fig:1}(a)), the propagation loss will
 be relatively small, \cite{ftn2,EM-textbook,VP_JOSA} due to small field penetration into the metal. What is however more important, is that the mode pattern in the wavegude that is calculated for the ``ideal metal'' boundary conditions,\cite{ftn3}  is identical to the standing wave formed by interfering plane waves incident on a planar dielectric layer, that does not at all involve any high-loss components -- see Fig. \ref{fig:1}(b). We will therefore use the waveguide geometry of Fig. \ref{fig:1}(a) as the way to clarify the underlying dynamics and the physical origin of the ghost waves, followed by the generalization of our approach to its lossless counterpart of Fig. \ref{fig:1}(b).
 
In the waveguide geometry of Fig. 1(a), the mode calculation is straightforward, and we obtain
\begin{eqnarray}
k_z & = & \pm \frac{1}{\sqrt{2}}  \left\{  \left(\epsilon_x + \epsilon_y\right) \left(\frac{\omega}{c}\right)^2 - \left(1 + \frac{\epsilon_x}{\epsilon_z}\right) q_x^2  \right. \nonumber \\
 & - & \left(1 + \frac{\epsilon_y}{\epsilon_z}\right) q_y^2
  \pm  \left[ \left( \left(\epsilon_x - \epsilon_y\right) \left(\frac{\omega}{c}\right)^2 \right. \right.
 \nonumber \\
 & + & \left. \left(1 - \frac{\epsilon_x}{\epsilon_z}\right) q_x^2 - \left(1 - \frac{\epsilon_y}{\epsilon_z}\right) q_y^2\right)^2 \nonumber \\
 & +  & \left. \left.  4 \left(1 - \frac{\epsilon_x}{\epsilon_z} \right)\left( 1 - \frac{\epsilon_y}{\epsilon_z} \right) q_x^2 \ q_y^2\ \right]^\frac{1}{2} \ 
 \right\}^{\frac{1}{2}},
 \label{eq:k_z}
\end{eqnarray}
where different signs correspond to different ``branches'' of the dispersion diagram, and  $\epsilon_x$, 
$\epsilon_y$ and $\epsilon_z$ are the primary components of the dielectric permittivity tensor in the waveguide core. 
For the waveguide  $q_x = m_x \pi/a_x$ and $q_y = m_y \pi/a_y$, where $a_x$ and $a_y$ represent the dimensions of the waveguide cross-section (see Fig. \ref{fig:1}(a)),  while in the 3D case of Fig. \ref{fig:1}(b) $q_x$ and $q_y$ correspond to the magnitudes of the in-plane momentum components of the incident wave(s).
If one of the integers $m_x$ and $m_y$ is equal to zero, the propagating fields are either TE- or TM-polarized,  otherwise the mode has the ``hybrid'' structure \cite{EM-textbook} when all six components of the electromagnetic field are nonzero.

\begin{figure*}[htbp] 
   \centering
    \includegraphics[width=6.75in]{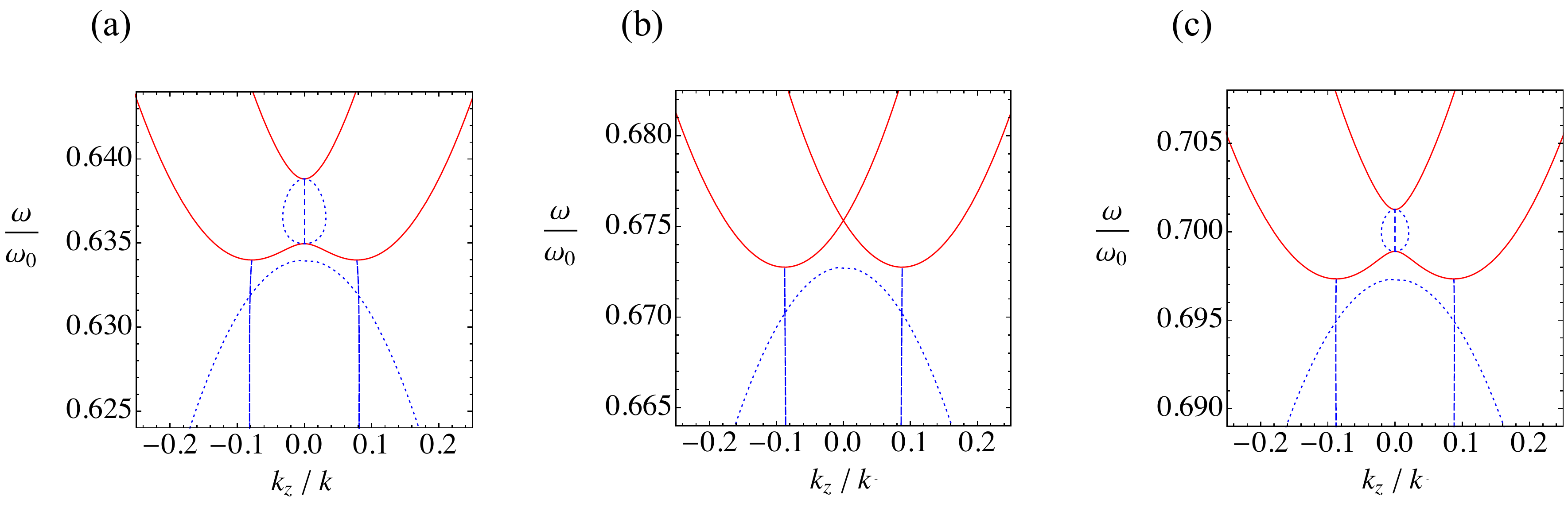}
       \caption{The dispersion diagrams for the waves supported by sodium nitrite ${\rm Na}{\rm NO_2}$, a biaxial anisotropic dielectric with the primary components of the dielectric permittivity tensor \cite{Moti_OE, Hirostu_JPSJ} $\epsilon_x \simeq 1.806$, $\epsilon_y \simeq 2.726$ and $\epsilon_z  \simeq 1.991$, with the frequency in units of $\omega_0$ and the wavenumber $k_z$ in units of $k \equiv \omega/c$,  for $q_x = 0.5 \ k_0$. $q_y = 0.75 \ k_0 $ (a), $q_x = 0.5 \ k_0$. $q_y \simeq 0.81 \ k_0$ (b),  and $q_x = 0.5 \ k_0$. $q_y = 0.85 \ k_0$ (c), where $k_0 \equiv \omega_0 /c$ is the free space wavenumber at the frequency $\omega_0$. For the waveguide system in Fig. \ref{fig:1}(a), $q_x = \pi m_x / a_x$ and $q_y = \pi m / a_y$, where $m_x$ and $m_y$ are positive integer numbers, while for the anisotropic  dielectric slab geometry  in Fig. \ref{fig:1}(b) $q_x$ and $q_y$ correspond to the magnitudes of the (in-plane) $x-$ and $y-$ components of the incident field wavevector. Note the Dirac point in (b) at the frequency $\omega \approx 0.675 \ \omega_0$. Red lines represent the propagating modes, while blue curves correspond to the ghost waves, with dashed and dotted lines showing the real and the imaginary parts of the wavenumber $k_z$. }
          \label{fig:2}
\end{figure*}

Choosing  $\hat{\bf x}$  as the direction of the largest permittivity in the $(x,y)$ plane (i.e. $\epsilon_x < \epsilon_y$), we find (see Fig. \ref{fig:2})  that the system supports negative index modes (with the group velocity $v_z = \partial\omega/\partial k_z < 0$) if and only if
\begin{eqnarray}
\epsilon_x < \epsilon_z < \epsilon_z,
\end{eqnarray}
and
\begin{eqnarray}
\arctan\left[ \frac{\epsilon_x}{\epsilon_y} \sqrt{\frac{\epsilon_y - \epsilon_z}{\epsilon_z - \epsilon_x}} \right]  < \theta < \arctan\left[ \sqrt{\frac{\epsilon_y - \epsilon_z}{\epsilon_z - \epsilon_x}} \right],
\end{eqnarray}
where the angle 
\begin{eqnarray}
\theta \equiv \arctan\left[q_y / q_x\right]
\end{eqnarray}
in the waveguide geometry  (Fig. \ref{fig:1}(a)) is defined by the dimensions of the waveguide cross-section ( \ $\theta = \arctan \left[m_y a_x / m_x a_y\right] $\ ),  while for  the 3D case (see Fig. \ref{fig:1}(b)) the angle $\theta$ corresponds to the propagation direction in the $(x,y)$ plane. 
This is consistent with the results of Ref. \cite{Moti_OE}, where the presence of negative index modes in biaxial anisotropic  media waveguides was first pointed out.

Furthermore, when $\theta$ is equal to 
\begin{eqnarray}
\theta_D & = & \arctan\left[\sqrt{ \frac{\epsilon_x}{\epsilon_y}  \cdot \frac{\epsilon_y - \epsilon_z}{\epsilon_z - \epsilon_x}} \ \right] , \label{eq:theta_D}
\end{eqnarray}
we find that the system shows Dirac dispersion point at the frequency
\begin{eqnarray}
\omega_D & = & c \ \sqrt{\frac{\left(\epsilon_y - \epsilon_x\right)}{\epsilon_y \left(\epsilon_z -\epsilon_x\right)}} \cdot q_x,
\label{eq:omega_D}
\end{eqnarray}
as seen in Fig. \ref{fig:2}(b).
Note that, as follows from Eqns. (\ref{eq:theta_D}) and (\ref{eq:omega_D}), the anisotropic waveguide system only shows the Dirac point when all three primary components of its dielectric permittivity tensor are different from each other.

For a given set of the integers $m_x$ and $m_y$ (or equivalently for a given magnitude of the in-plane momentum on the incident field $(q_x, q_y)$ (see Fig. \ref{fig:1}(a)), the biaxial anisotropic dielectric core supports propagating waves only above the critical frequency (see Fig. \ref{fig:2})
\begin{eqnarray}
\omega_c & = &  \frac{c}{\sqrt{\epsilon_y - \epsilon_x}} \left(q_x \ \sqrt{\frac{\epsilon_z}{\epsilon_x} - 1} + q_y \ \sqrt{\frac{\epsilon_z}{\epsilon_x} - 1} \  \right), \label{eq:omega_c} 
\end{eqnarray}
when two simultaneous tangent bifurcations, one at  $k_z = k_c$ and the other at $k_z = - k_c$, where 
\begin{eqnarray}
k_c & = &  \frac{1}{\sqrt{\epsilon_z \left(\epsilon_y - \epsilon_x\right)}} 
\left\{ 
\epsilon_x \left(\epsilon_z - \epsilon_y\right) q_x^2 + \epsilon_y \left(\epsilon_x - \epsilon_z\right) q_y^2 \right.\nonumber \\
& + & 
\left.
 \left(\epsilon_x + \epsilon_y\right) \sqrt{\left(\epsilon_z - \epsilon_x\right)\left(\epsilon_y -\epsilon_z\right)} q_x q_y
\right\}^{1/2},
\end{eqnarray}
create two pairs of positive and negative-index modes. However, if -- following the standard approach originally developed for the semiclassical dynamics of non-integrable systems,\cite{Kus,ghost2,ghost3,ghost4}  -- one extends the system phase space  to the complex domain, Eqn. (\ref{eq:k_z}) also yields two pairs of solutions below the critical frequency $\omega_c$, with complex-conjugate wavenumbers that have simultaneously nonzero real and imaginary parts : $\pm k_z' \pm i k_z''$, as shown in Fig. \ref{fig:2} by blue lines. These new modes are the optical equivalents to the ``ghost'' solutions in non-linear dynamics.\cite{Kus}

\begin{figure}[htbp] 
   \centering
    \includegraphics[width=2.95in]{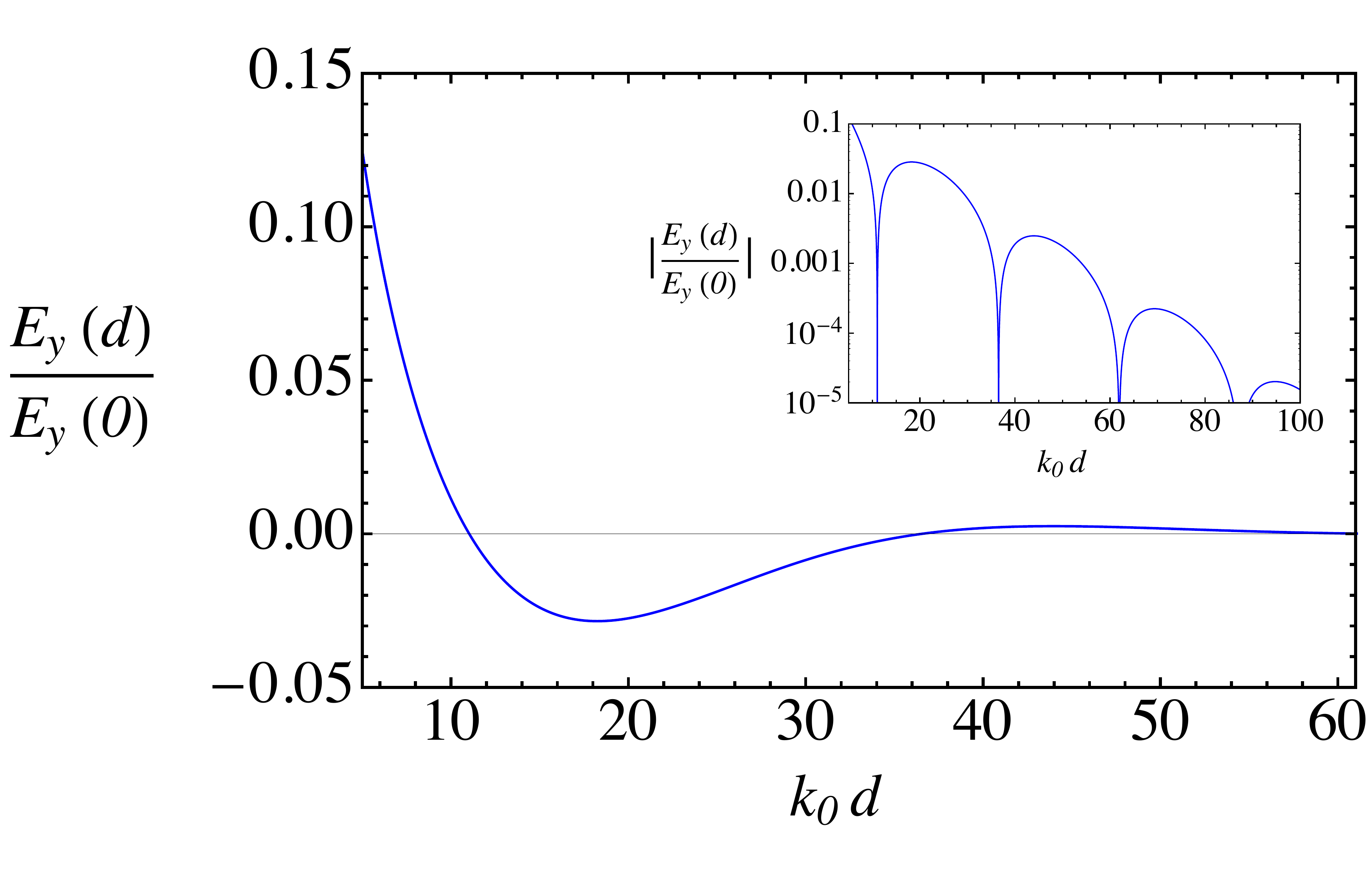}
     \caption{The $y$-component of the electric field in the biaxial anisotropic dielectric slab as a function of the slab width $d$. The dielectric occupies the range $0 < z < d$ (see also the schematics in Fig. \ref{fig:4}(a,b)), $q_x  \simeq 0.79 \ k_0$ and $q_y \simeq 0.185 \ k_0$. The surrounding medium is air, and the anisotropic dielectric is sodium nitrite ${\rm Na}{\rm NO_2}$. The inset shows the field in the logarithmic scale. Note the combination of the exponential decay and the oscillatory behavior, characteristic of the ghost waves. }     
     \label{fig:3}
\end{figure}

\begin{figure*}[htbp] 
   \centering
    \includegraphics[width=6.9in]{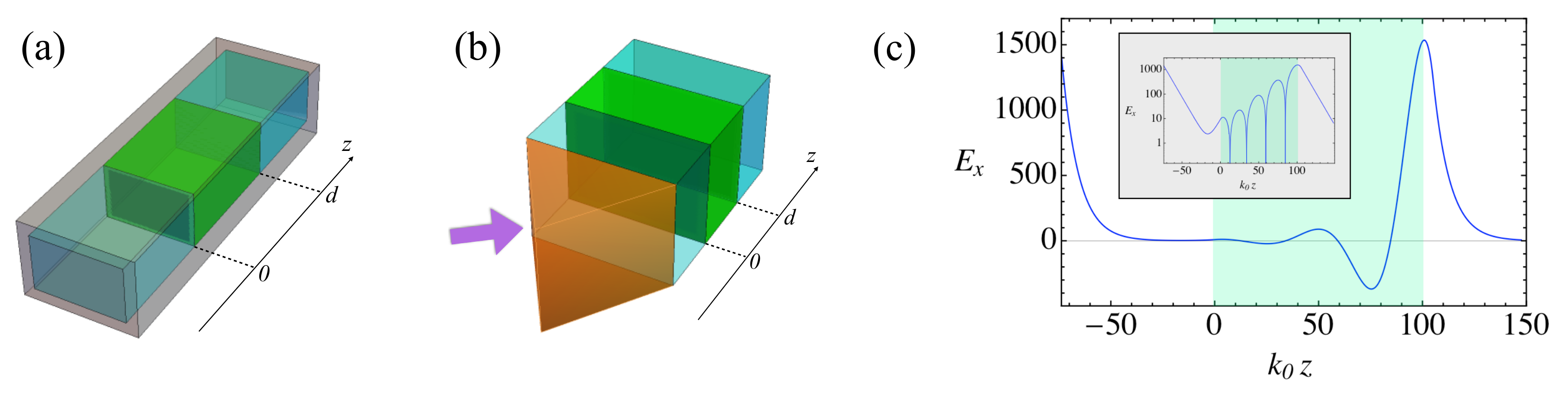}
     \caption{Evanescent field enhancement at the ghost resonance, in the waveguide (a) and in slab geometry (b). Green region represents the biaxial anisotropic dielectric, blue is the surrounding dielectric medium, and orange is the high-index prism coupler. Panel (c) shows the $x$-component of the electric field, with the linear scale in the main plot and the logarithmic scale in the inset. The green-shaded areas in panel (c) and its inset indicate the range occupied by the anisotropic dielectric medium. In this example, the anisotropic dielectric is  sodium nitrite ${\rm Na}{\rm NO_2}$ with the width $d \simeq 16.18 \ \lambda_0$, and the surrounding medium is sugar solution with the permittivity of $\epsilon_0 = 2.01$, $q_x \simeq 0.793 \ k_0$ and $q_y \simeq 1.182 \ k_0$. Note the dramatic enhancement of the incident evanescent field in the anisotropic material.}     
     \label{fig:4}
\end{figure*}

Qualitatively, the ghost fields combine the properties of the evanescent and propagating waves. Similarly to the evanescent  modes, the intensity of the ghosts waves exponentially decays with distance. On the other hand, the ghost waves also oscillate -- which allows to use conventional interference to control their propagation. The latter behavior is illustrated in Fig. \ref{fig:3}, corresponding to a
 biaxial dielectric of length $d$ inserted into the originally air-field metal waveguide (see Fig. \ref{fig:4}(a)  ) that is operating below the cut-off frequency $\omega_c$, or to a biaxial dielectric slab of the width $d$ (see Fig. \ref{fig:4}(b)). As a function of the length $d$, the field in the dielectric shows both the exponential decay and the oscillations, consistent with the picture of the excitation of the ghost waves.

The most important feature of the ghost waves, however, is that they can be resonantly coupled to the incident evanescent wave. In this regime, within the biaxial dielectric supporting the ghost waves, the evanescent decay is replaced by the exponential increase of the field amplitude -- as seen in Fig. \ref{fig:4}. Aside from the oscillatory behavior of the ghost waves ``under'' the exponential envelope, this behavior is similar to that in the negative index superlens that also shows exponential ``amplification'' of the evanescent field. \cite{Pendry} However, as the ghost waves can be supported by a dielectric slab that does not include any lossy components such as metals (see Fig. \ref{fig:4}(b)), the ghost waves are free from the constraint of the material absorption that severely limits the superlens performance.

\begin{figure}[htbp] 
   \centering
    \includegraphics[width=2.9in]{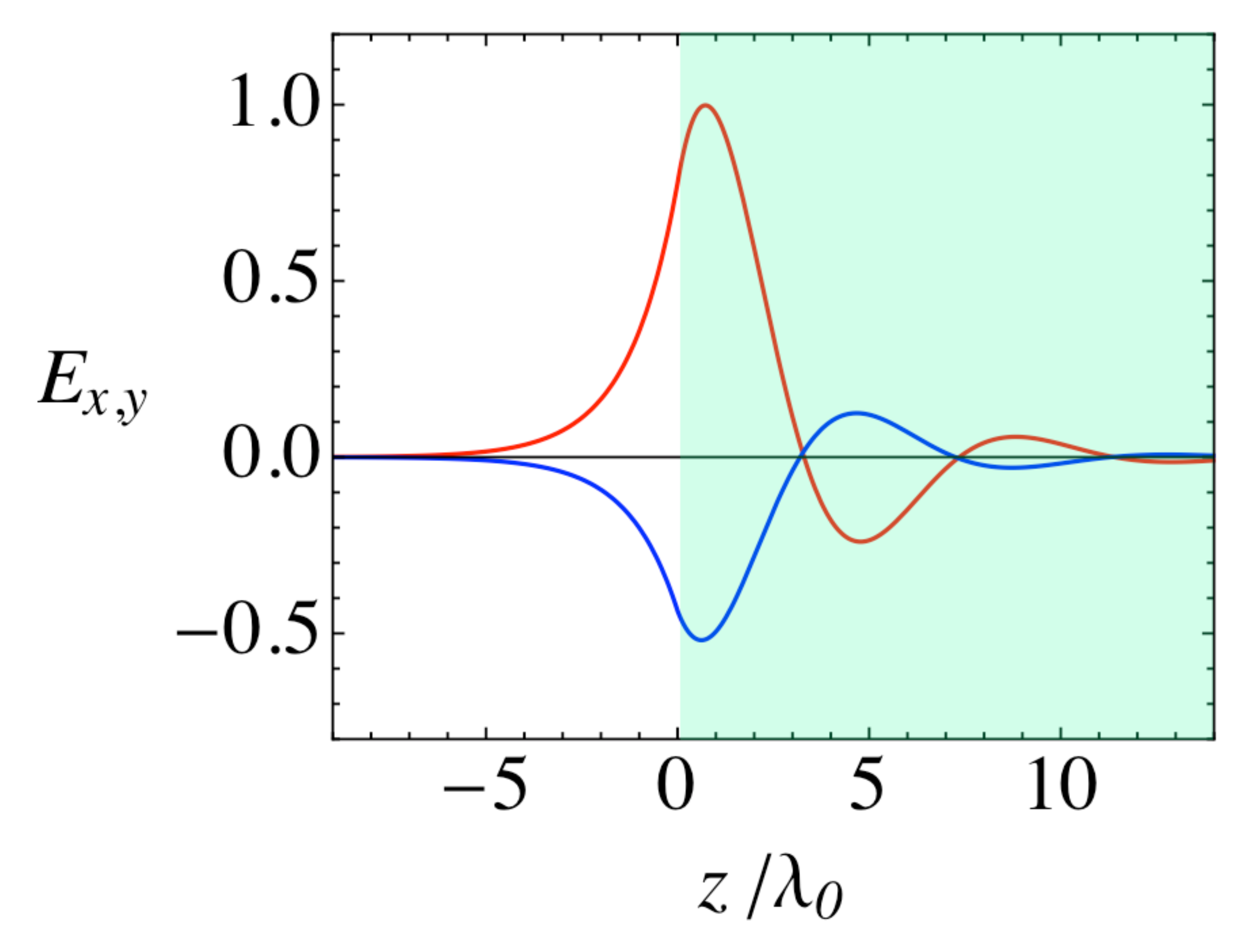}
     \caption{The surface state profile at the interface of isotropic dielectric with biaxial anisotropic medium. The red and blue lines  respectively correspond to the $x-$ and $y-$ components of the electric field. The biaxial dielectric is sodium nitrite ${\rm Na}{\rm NO_2}$ ( $\epsilon_x \simeq 1.806$, $\epsilon_y \simeq 2.726$, $\epsilon_z  \simeq 1.991$), the isotropic medium is sugar solution ($\epsilon = 2.01$), $q_x \simeq 0.793 \ k_0 $ and $q_y \simeq 1.182 \ k_0$. Note the simple exponential decay in the isotropic material and oscillatory behavior in the biaxial medium. }
          \label{fig:5}
\end{figure}

The fundamental physical origin of the exponential ``amplification" in biaxial media is also similar to that in the negative index superlens, -- it's the resonant coupling to surface states supported by the medium.\cite{Haldane_arxiv} In the case of the negative index material  - air interface, the surface state in question is the surface plasmon-polariton,\cite{Haldane_arxiv, Vlad-book} while in the biaxial dielectric it's the surface mode at the isotropic dielectric - anisotropic  dielectric interface, shown in Fig. \ref{fig:5}. This surface modes is formed by the ``regular" evanescent field in the isotropic side, and by the decaying ghost wave in the biaxial medium (note the oscillatory behavior for $z > 0$ in Fig. \ref{fig:5}). In the limit $\epsilon_z \to \epsilon_x$ this mode reduces to the well-known Dyakonov surface wave that was originally introduced for the interface of an isoroptic medium with a uniaxial dielectric.\cite{Dyakonov,Dyakonov_extended,Dyakonov_review}

At the interface of the biaxial medium and the isotropic dielectric with the permittivity $\epsilon_0$, the surface state dispersion $\omega_s\left(q_x,q_y\right)$ is defined by the equation
\begin{eqnarray}
& & \kappa_0 \left(\kappa_+ + \kappa_-\right) \left\{ \kappa_+ \kappa_- + \frac{\epsilon_x \epsilon_y}{\epsilon_0} \left(\frac{q_x^2}{\epsilon_y} + \frac{q_y^2}{\epsilon_x} - \left(\frac{\omega_s}{c}\right)^2 \right) \right\}
\nonumber \\
& + & \kappa_+ \kappa_- \left\{ \left(1 + \frac{\epsilon_x}{\epsilon_0}\right) q_x^2 + \left(1 + \frac{\epsilon_y}{\epsilon_0}\right) q_y^2 
\right. \nonumber \\
& - & \left. \left(\epsilon_x + \epsilon_y\right) \left(\frac{\omega_s}{c}\right)^2 \right\} \nonumber \\
& + & \left\{ \kappa_+^2  \kappa_-^2 + \frac{\epsilon_x \epsilon_y}{\epsilon_0} \kappa_0^2   \left(\frac{q_x^2}{\epsilon_y} + \frac{q_y^2}{\epsilon_x} - \left(\frac{\omega_s}{c}\right)^2 \right) \right\} = 0,
\label{eq:surfstate}
\end{eqnarray}
where
\begin{eqnarray}
\kappa_0 & = & \sqrt{q_x^2 + q_y^2 - \epsilon_0 \left(\frac{\omega_s}{c}\right)^2},
\end{eqnarray}
and
\begin{eqnarray}
\kappa_\pm & = & \frac{1}{\sqrt{2}} \left\{   \left(1 + \frac{\epsilon_x}{\epsilon_z}\right) q_x^2
+ \left(1 + \frac{\epsilon_y}{\epsilon_z}\right) q_y^2 \right. \nonumber \\
& - &  \left(\epsilon_x + \epsilon_y\right) \left(\frac{\omega_s}{c}\right)^2  
\pm \left[ \left( \left(\epsilon_x - \epsilon_y\right) \left(\frac{\omega_s}{c}\right)^2 \right. \right.
  \nonumber \\
 & + & 
  \left(1 - \frac{\epsilon_x}{\epsilon_z}\right) q_x^2
 - \left.  \left(1 - \frac{\epsilon_y}{\epsilon_z}\right) q_y^2\right)^2 \nonumber \\
 & + & \left. \left.  4 \left(1 - \frac{\epsilon_x}{\epsilon_z} \right)\left( 1 - \frac{\epsilon_y}{\epsilon_z} \right) q_x^2 \ q_y^2\ \right]^\frac{1}{2} \ 
 \right\}^{\frac{1}{2}}.
\end{eqnarray}
Note that Eqn. (\ref{eq:surfstate}) has a solution only when $\epsilon_z < \epsilon_0 < \epsilon_y$. 

\begin{figure}[htbp] 
   \centering
    \includegraphics[width=2.9in]{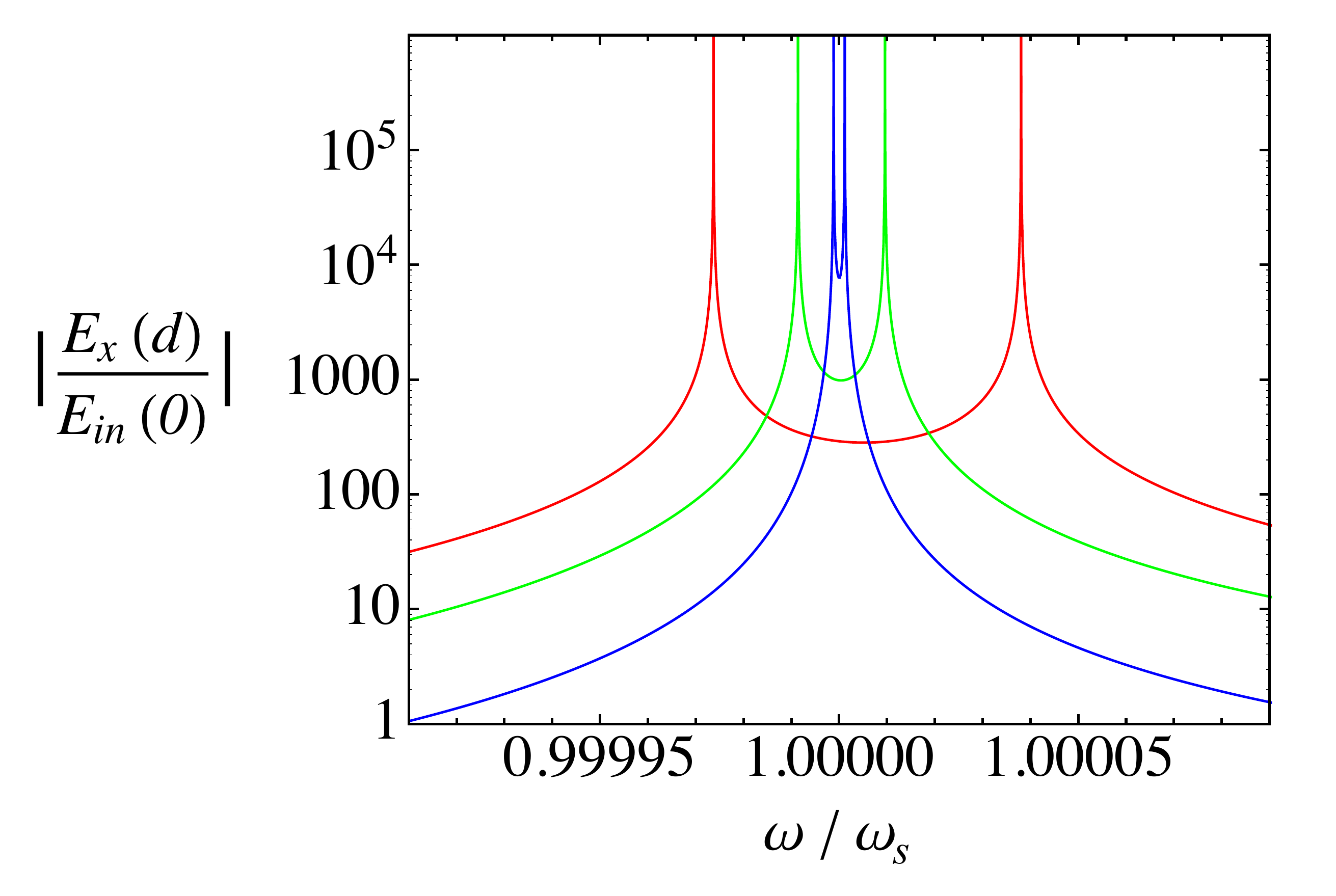}
        \caption{The frequency spectrum  of the ``transmitted'' field $E_x(d)$, normalized
        to the amplitude of the incident field $E_{\rm in}$ at $z = 0$ (see the schematics in Fig. \ref{fig:4}(a),(b)),
        for the sodium nitrite biaxial crystal slab (green region in  Fig. \ref{fig:4}(a),(b)), surrounded by sugar 	   
        solution with the permittivity $\epsilon = 2.01$ (shown in blue  in Fig. \ref{fig:4}(a),(b)), with $q_x \simeq 0.79 \  k_0$ and $q_y \simeq 1.18 \ k_0$. The width of the dielectric slab $ d \simeq  7.96 \ \lambda_0 $ (red curve), $d \simeq 11.94 \  \lambda_0$ (green curve) and $ d \simeq 16.78 \ \lambda_0$ (blue curve). The frequency is normalized to the value corresponding to the surface state at a single sodium nitrite - sugar solution interface, given by Eqn. (\ref{eq:surfstate}).  }
          \label{fig:6}
\end{figure}

As a function of frequency, the transmission coefficient for the incident evanescent field shows a very sharp resonance which becomes progressively more narrow with the increase of the width of the anisotropic layer $d$ -- see Fig. \ref{fig:6}. Note that close to the center of this ``ghost resonance'' its profile shows the double-peak structure, corresponding to the symmetric and anti-symmetric combination of the surface states at the two interfaces. An increase of the width of the anisotropic layer reduces the coupling between these two surface states, which reduces the splitting between the peaks -- see Fig. \ref{fig:6}.

While for the idealized model of the system (lossless dielectric, perfect planar interfaces, no bulk defects leading to light scattering), the transmission coefficient at the ghost resonance can be arbitrary high, any disorder will
 limit the field enhancement. While these imperfection and their impact can  be, at least in theory, reduced to an arbitrarily small level, these is also a fundamental limit to the performance of the proposed ``ghost resonator,'' imposed by  the inherent non-locality of the electromagnetic response due to the final size of the material unit cell. For natural dielectric media, the corresponding correction to the permittivity scales \cite{LL_cont_media}  as $(k_0 a_0)^2$, where $a_0$ is on the order of the atomic/molecular size, leading to the fundamental limit on the evanescent field enhancement  $\sim 1/ \left(k_0 a_0\right)^2 \sim  10^5$.

In conclusion, we have demonstrated that biaxial anisotropic media supports ghost waves that combine the properties of the propagating and the evanescent fields. We have shown that resonant coupling of the ghost modes can be used to exponentially enhance and modulate the incident evanescent waves, thus opening a new way of the near field control and manipulation. 

This work was partially supported by the National Science Foundation (grant 1629276-DMR), Army Research Office (grant W911NF-14-1-0639) and Gordon and Betty Moore Foundation.


\begin{thebibliography}{1}
\bibitem{ftn1}
{The situation is similar in polaritonic materials in mid-IR wavelength range, where the negative permittivity originated from resonant coupling to optical phonons, with the concomitant ``resonant" loss. \cite{Vlad-book}}
\end{thebibliography}



\end{document}